\documentclass[aps,twocolumn,epsfig,rotate,showpacs,floatfix,roupedaddress,superscriptaddress]{revtex4}
\usepackage{epsfig} 
\usepackage{graphicx} 
\usepackage{amsmath} 
 
\usepackage{graphicx} 
\begin{document} 

\title{A Simple Three-Parameter Model Potential For Diatomic 
Systems:  From Weakly and Strongly Bound Molecules to 
Metastable Molecular Ions}  
 
\author{Rui-Hua Xie} 
 
\affiliation{Department of Physics and Institute for Quantum 
Studies, Texas A\&M University, TX 77843} 
 
\author{Jiangbin  Gong} 
 
\affiliation{Department of Chemistry and James Franck Institute,  
University of Chicago, Chicago, IL 60637} 
 
\begin{abstract} 
Based on a simplest molecular orbital theory of H$_{2}^{+}$,  
a three-parameter model potential function is proposed to  
describe ground-state diatomic systems with closed-shell and/or 
S-type valence-shell constituents over a significantly wide range 
of internuclear distances.  More than 200  weakly and  strongly bound 
diatomics have been studied, including neutral and  singly-charged 
diatomics (e.g., H$_{2}$, Li$_{2}$, LiH, Cd$_{2}$, Na$_{2}^{+}$, 
and  RbH$^{-}$), long-range bound diatomics (e.g.,  NaAr, CdNe, 
He$_{2}$, CaHe, SrHe, and BaHe), metastable molecular dications 
(e.g., BeH$^{++}$, AlH$^{++}$,  Mg$_{2}^{++}$, and LiBa$^{++}$), 
and molecular trications (e.g., YHe$^{+++}$ and ScHe$^{+++}$). 
 
 
\end{abstract} 
 
\maketitle

Modeling the interaction potential of diatomic systems is of fundamental  
importance to many issues~\cite{g00,w81,kop03a,p01,t95,k99,e99,doye05},  
including  atom-atom collisions, molecular spectroscopy, prediction of 
cluster structures,  molecular dynamics simulation, chemical reactivity,  
matter-wave interferometry,  and transport properties for more complex  
systems. Also of great interest are the potential functions for long-lived 
metastable doubly- or multiply-charged ions~\cite{mathur}  that are 
relevant to high-density energy storage materials and  to  
characterization and analytical methods for biosystems. 
 
Modern spectroscopy, diffraction, and scattering techniques~\cite{g00,kop03a}   
provide a direct experimental approach to  studies of interaction potentials  
of diatomic systems. In particular, diatomic potentials can be inferred from  
the spectroscopy data by three general approaches~\cite{kop02a}:  
(i) the Wentzel-Kramers-Brillouin (WKB) Rydberg-Klein-Rees (RKR) method,  
(ii) the WKB-based Dunham approach, and  
(iii)  semiempirical or empirical procedures.  
On the theoretical side, a diatomic  potential curve may be predicted  
directly by {\sl ab initio}  calculations~\cite{kolos65b} and quantum Monte 
Carlo simulations~\cite{alex04a}.  These theoretical methods can, in principle, 
be very accurate when sufficient  electronic configurations are included 
in the calculations, but can be  prohibitively expensive in weakly bound 
systems\cite{w81} and/or many-electron systems\cite{kop03a}.  
 
Numerous attempts to analytically model diatomic potentials have been  
made~\cite{kop03a,t95,kop02a,zav91a,cve94,bel02,cah04}. The well-known  
potential functions include Morse, Born-Mayer, Hulburt-Hirschfelder,   
Rosen-Morse, Rydberg,  P\"{o}schl-Teller, Linnett, Frost-Musulin, Varshni III,  
Lippincott, Lennard-Jones, and Maitland-Smith potentials~\cite{kop03a,kop02a},  
as well as the celebrated Tang-Toennies potential~\cite{t95} and  the recently  
proposed  Morse-based potentials \cite{zav91a}. These potentials usually aim  
to describe either strongly or weakly bound, neutral or singly-charged  
diatomics and often lose their validity for either small or relatively large  
internuclear distance (denoted  $R$ hereafter). Thus, recent effort has been 
devoted  to the construction of hybrid potentials, which use different 
functions for  different  interaction regions of 
$R$~\cite{kop03a,kop02a,cve94,bel02,cah04}  
and thereby  need more than four potential parameters. Well-known examples of  
hybrid potentials include the combined Morse-van der Waals~\cite{kop02a},  
general Buckingham-type exp(n,m)~\cite{kop02a}, Cvetko~\cite{cve94}, and  
Bellert-Breckenridge~\cite{bel02} potentials, as well as the  most recently  
proposed Rydberg-London  potential~\cite{cah04}. For metastable doubly- or  
multiply-charged molecules, none of the above-mentioned potential functions  
is able to describe their ground states. To date, only few theoretical
models\cite{gill88,zhu} that were specifically designed for metastable molecular
 dications~\cite{mathur}  have been proposed.
 
The goal of this Letter is twofold.  First, we propose a molecular-orbital  
theory based approach to obtain  a very simple analytical potential of 
diatomic  systems. The potential function thus obtained has significant 
applicability  insofar as it can describe a wide variety of diatomic 
molecules (including both weakly and strongly bound systems)  with good 
accuracy for almost the whole range of $R$ but excluding  the large-$R$ 
limit. Second, we show that this potential function can also describe 
metastable doubly-charged diatomics as well as singly- and triply-charged 
ones. Specifically,  we advocate a very simple three-parameter  
ground-state  potential function that  is applicable to  more than 
200 diatomics with  closed-shell and/or S-type valence-shell  
constituents (atoms or ions whose shells are closed or whose valence shells 
are S-orbital). These include  neutral and singly-charged diatomics, 
long-range bound diatomics~\cite{stwalley},  metastable molecular 
dications~\cite{ornellas93a} and  molecular  trications~\cite{wesen}. The
 details  for these systems  and the  associated  parameters 
of our model potential are given in Ref.\cite{supp}. 
 
We require a few-parameter potential function to satisfy the following basic  
conditions:  
  (i) Its asymptotic value $E_{\infty}$  for   $R\rightarrow\infty$ is finite.  
 (ii) A global potential minimum $E_{\min}$ at the equilibrium distance  
      R$_{e}$ is allowed.  
(iii) It approaches infinity as $R\rightarrow 0$.  
(iv) One local potential maximum $E_{\max}$ at $R_{\max}$ is  allowed  
     to describe metastable systems.  
 (v) Both Coulomb and exchange interactions can be described by using only few 
     parameters. 
To seek such a potential function we revisit the  molecular-orbital 
theory~\cite{mcquarrie} as applied to H$_{2}^{+}$, the simplest 
single-electron diatomic system, with the associated Hamiltonian 
$H = -\frac{1}{2}\bigtriangledown^{2}
     -\frac{1}{r_{A}}-\frac{1}{r_{B}}+\frac{1}{R}$ (in atomic units),
where $r_{A}$ and $r_{B}$ denote  electron-nucleus distances. 
This case can be solved exactly, but here it is used as a 
reference system to understand  how  the simplest version of the  
molecular-orbital theory may be improved. To  that end consider the 
S-type trial function of  H$_{2}^{+}$: 
$\Psi =c_{1}|\phi_{0}^{A}\rangle + c_{2}|\phi_{0}^{B}\rangle$,  
where $|\phi_{0}\rangle=\frac{e^{-r}}{\sqrt{\pi}}$ (the 1s-orbital of H 
atom). The energy of the bonding orbital is then given by  
\begin{equation} 
E(R)=E_{\infty} + \frac{J_{1}(R)+K_{1}(R)}{1+S_{0}(R)},  
\end{equation} 
where $E_{\infty}=-\frac{1}{2}$, 
$J_{1}(R) = e^{-2R}\left(1+\frac{1}{R}\right)$,  
$K_{1}(R) = e^{-R}\left(\frac{1}{R}-\frac{2}{3}R\right)$, and  
$S_{0}(R) = e^{-R}\left(1+R+\frac{1}{3}R^{2}\right)$ ~\cite{supp,mcquarrie}.  
In the literature~\cite{t95,mcquarrie},  $J_{1}$ and $K_{1}$ are called 
the Coulomb and exchange integrals, respectively, and $S_{0}$ is the 
overlap integral  between the orbitals 
$|\phi_{0}^{A}\rangle$ and $|\phi_{0}^{B}\rangle$. 
Figure 1(a) shows the resultant potential curve of H$_{2}^{+}$. The minimum 
energy  E$_{min}$ is -0.56483 hartree, located at $R_{e}=2.500$ bohr. This 
should  be compared with the most accurate data~\cite{peek65a}: 
E$_{\min}=-0.60263$  hartree, at R$_{e}=1.999$ bohr. Clearly then, while 
the  analytical potential function  of H$_{2}^{+}$ derived above  satisfies 
most of  the general pair potential requirements set above, quantitatively 
it  should be improved. Indeed, if  {\sl polarization} and even {\sl diffuse} 
functions  are included in the trial function, then the potential curve in 
the bonding region has a  much better performance. As illustrated in 
Fig.~1(b),  by  using couple cluster method with single and double excitation  
(CCSD)~\cite{ccsd}  with  STO-3G (1s orbital only),  6-31G(d,p) (including 
polarization function), and   6-311++G(3df,3pd) (including diffuse functions) 
Gaussian-type basis sets, one obtains 
$E_{\min}=-0.582697, -0.594490,  -0.602207$ 
hartree at $R_{e}=2.004280,  1.948160, 1.999899$ bohr, respectively.   

We now introduce a simple analytical potential function to improve the above 
potential for H$_{2}^{+}$. That is, 
\begin{equation} 
E(R,\alpha,\beta,\gamma)=E_{\infty} +
 \frac{J_{1}(R,\gamma)+K_{1}(R,\alpha,\beta)}{1+ S_{0}(R)},
\end{equation}
where parameter $\gamma$ is introduced in the Coulomb integral $J_{1}$, 
i.e., $J_{1}(R,\gamma) = e^{-2\gamma R}\left(1+\frac{1}{R}\right)$, 
and two parameters $\alpha$ and $\beta$ are introduced in
the exchange integral $K_{1}$, 
i.e., $K_{1}(R,\alpha,\beta) = e^{-\alpha R}
\left(\frac{1}{R}-\beta R\right)$. 
Below we briefly discuss the meanings of the three parameters in the 
light of the polarization approximation\cite{t95}. A detailed 
discussion of this issue is presented in Ref.\cite{supp}. In 
the first-order polarization approximation, Eq. (2) 
can be rewritten as $E(R,\alpha,\beta,\gamma) = E(E_{p}, \epsilon_{ex}) 
=E_{p} - [1-S_{0}(R)]\epsilon_{ex}$, where $E_{p}=E_{\infty}+J_{1}(R,\gamma)$ 
and  $\epsilon_{ex} =\frac{S_{0}(R)}{1-S_{0}(R)^{2}}
[J_{1}(R,\gamma)-K_{1}(R,\alpha,\beta)]$ 
are the polarization and exchange energies, respectively 
(For one-electron H$_{2}^{+}$, the {\sl exchange energy}
can be interpreted as resulting from the electron hopping back and
forth across the median plane between two protons~\cite{t95}, therefore
refering to the exchange of two protons). Clearly, parameter $\gamma$ 
directly adjusts $J_{1}(R,\gamma)$  and hence the polarization 
energy $E_{p}$. Because $\epsilon_{ex}$
also depends on $J_{1}(R,\gamma)$, the introduction of
$\gamma$ also affects the dispersion (positive) part of
$\epsilon_{ex}$.  Through the term $K_{1}(R,\alpha,\beta)$,
parameters $\alpha$ and $\beta$
are used to account for the $R$-dependence of 
$\epsilon_{ex}$ that is already affected by $\gamma$.
In particular, the {\sl induction} part (the negative term) 
of $\epsilon_{ex}$ is adjusted  only 
by  parameter $\alpha$,  and parameter $\beta$ further adjusts 
the dispersion part of $\epsilon_{ex}$ through
the negative term of $K_{1}(R,\alpha,\beta)$.
Certainly there are alternative approaches for realizing
these adjustments, but the new potential function constructed above
includes  both the  Pauli repulsive term $\frac{e^{-b R}}{R}$ 
and the well-known Born-Mayer ``exponential"
form  $Ae^{-bR}$. This is different from Tang-Toennis~\cite{t95},
Cvetko~\cite{cve94}, and Rydberg-London~\cite{cah04} potentials,
whereas only the Born-Mayer form appears as their repulsion terms.
It should also be stressed that although  $E(R,\alpha,\beta,\gamma)$ 
now has three adjusting parameters, it is still analogous to Eq.~(1) in many 
aspects (e.g., satisfying all the  pair potential requirements set above). 
Based on this three-parameter potential function,  we find that 
the potential curve for H$_{2}^{+}$, as shown in Fig. 1(a), 
would agree very  well with the most accurate data available in the  
literature~\cite{peek65a} if we choose $\alpha=1.0511106$, 
$\beta=0.917034242$ and  $\gamma=2.25$. This confirms that  
$\alpha $, $\beta $, $\gamma $ can be properly adjusted such that 
contributions of both  the polarization and exchange energies can 
be accounted for in an efficient way, thereby achieving, in effect, the 
same goal as that of using larger basis sets (polarization, diffuse 
functions) in the trial wavefunctions.

\begin{figure}
\epsfig{file=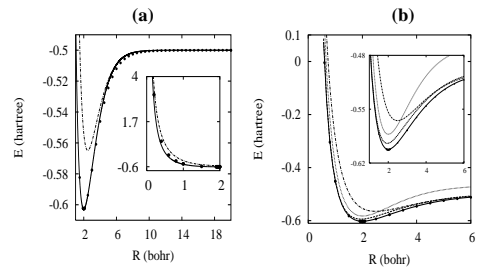, width=7cm}
\caption{\label{fig:fig1} \small
The potential energy curve of the ground state of H$_{2}^{+}$:
(a) Equation (1) (dot-dashed line) and Equation (2) (solid line,
$\alpha=1.0511106$, $\beta=0.917034242$, $\gamma=2.25$ ); (b) Equation
(1) (dot-dashed line), CCD/STO-3G (dotted line), CCSD/6-31G(d,p)
(dashed line), and  CCSD/6-311++G(3df,3pd) (solid line). The filled dots
in (a) and (b) are the most accurate data reported in the
literature~\cite{peek65a}. The inset in (a) is for the short-range region.}
\end{figure}

\begin{figure}
\epsfig{file=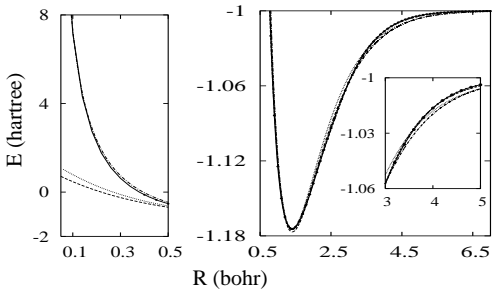, width=7cm}
\caption{\label{fig:fig2}\small
 The comparison between the new potential
(dot-dashed line, $\alpha=1.5065756$, $\beta=2.48475652$,
$\gamma=1.45$), 11-parameter-fit model potential (solid line,
Ref.~\cite{agu94}), hybrid Rydberg-London potential
(dashed line, Ref.~\cite{cah04}),  Morse potential (dotted line)
and the most accurate \textit{ab initio} data (filled circles,
Ref.~\cite{kolos65b}) for hydrogen molecule H$_{2}$. Inset in the
rightmost figure is the enlarged part between 3.0 and 5.0 bohr.}
\end{figure}

Certainly our real motivation is to extend this simple and successful 
procedure from H$_{2}^{+}$ to other multi-electron diatomic systems. 
A number of established  results about the electronic structures of 
diatomic systems suggest that this is possible  for ground-state diatomics 
with closed-shell and/or S-type constituents  (see details 
in Ref.\cite{supp}).  In particular, in the zeroth-order approximation,
the outermost electrons in a multielectron system move in the Hartree-Fock 
self-consistent field or the effective potential of all the core electrons 
and the positive nucleus, and the asymptotic exchange energy of a multielectron 
system can arise primarily from the outermost electrons. The exchange 
interactions between two multielectron atoms, which play a crucial role in 
chemical bonding, are dominated by the exchange of a single pair of electrons, 
and the associated exchange energy is given by that of a single electron pair 
multiplied by a constant.  Based on the polarization approximation~\cite{t95}, 
the ground-state potentials $E(E_{p},\epsilon_{ex})$ 
of H$_{2}$ and other multielectron diatomic  systems, when 
expressed in terms of the polarization and exchange energies, 
can take a similar form~\cite{supp} to that of H$_{2}^{+}$, 
despite that their origins of the exchange energy are totally 
different.  Motivated by these known theoretical results, we 
have carried out extensive studies of  more than 200 diatomic systems for 
which experimental or \textit{ab initio} data  are available. We find that, 
indeed, the above three-parameter potential can be proposed as a widely 
applicable potential function for ground-state diatomics with closed-shell 
and/or S-type constituents.

\begin{figure}
\epsfig{file=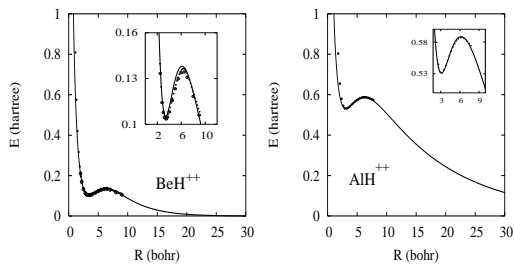,width=7cm}
\caption{\label{fig:fig3}
\small The potential energy curve of the ground state of BeH$^{++}$
($\alpha=0.687$, $\beta=1.43632004$, and $\gamma=0.1185$) and AlH$^{++}$
($\alpha=0.585984$, $\beta=0.796691521$, and $\gamma=0.0365$). The filled/open
circles denote the previous data from Ref.~\cite{ornellas93a}.}
\end{figure}

\begin{figure}
\epsfig{file=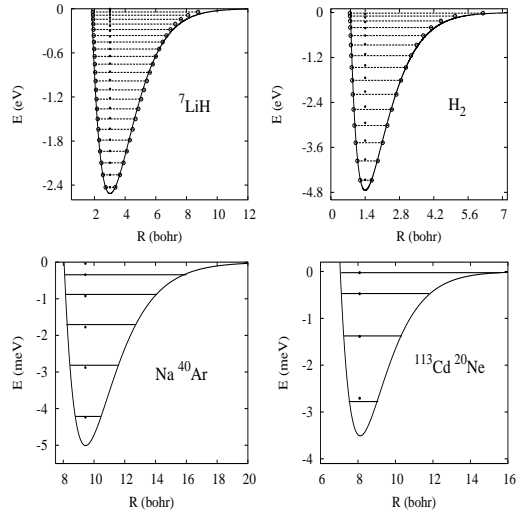,width=7cm}
\caption{\label{fig:fig4} \small 
 The computed  vibrational levels (filled circles)
of  $^{7}$LiH ( $\alpha=0.8885591$, $\beta=1.51479003$, $\gamma=0.345$),
H$_{2}$,  Na$^{40}$Ar ($\alpha=0.6365$, $\beta=0.0229701414$,$\gamma=0.50$),
and  $^{113}$Cd$^{20}$Ne ( $\alpha=0.89314$, $\beta=0.33243839$,
$\gamma=0.40$). The solid horizontal lines denote the measured vibrational
levels for Na$^{40}$Ar~\cite{smalley79a} and
CdNe (suggested values from Ref.~\cite{kop02b} based on the average mass
of CdNe). The open circles (dashed lines) are RKR potential points (measured
vibrational levels) for H$_{2}$~\cite{tobias61a} and $^{7}$LiH~\cite{chan86a}.}
\end{figure}

To determine the three parameters of the proposed potential function
we suggest several numerical approaches in the Appendix C of our
supplementary material\cite{supp}. The model potential curves
thus determined for  more than 200 weakly and strongly
bound diatomic systems agree with the available experimental or theoretical 
data, with the agreement  in many cases much better than one could 
naively anticipate from a  three-parameter potential (see Table~1 and 
Fig.~A in Ref.\cite{supp}). Below we discuss some sample results. In 
particular,  Fig.~2 shows that the potential curve for H$_{2}$ 
is in good agreement with  the recent 11-parameter model 
potential~\cite{agu94} and the most accurate 
data~\cite{kolos65b}, thereby giving a better performance than the Morse  
function\cite{kop03a} and the most recent hybrid Rydberg-London 
potential~\cite{cah04}.  Even more significantly, our potential 
function is applicable to metastable S-type  molecular 
dications~\cite{ornellas93a} (e.g.,  He$_{2}^{++}$, Be$_{2}^{++}$,  
BeH$^{++}$, Mg$_{2}^{++}$, MgH$^{++}$, BH$^{++}$, AlH$^{++}$, 
LiBa$^{++}$,  KBa$^{++}$, NaBa$^{++}$, and Ba$_{2}^{++}$), and  molecular  
trications~\cite{wesen} (e.g.,YHe$^{+++}$, ScHe$^{+++}$) as well as neutral  
and singly-charged diatomic systems. The potential curves for BeH$^{++}$ and  
AlH$^{++}$ using our potential function are shown in Fig.~3,  where the 
potential barriers agree well with the  literature data\cite{ornellas93a}. 
Figure~4 displays the calculated rotationless  vibrational 
levels for $^{7}$LiH, H$_{2}$, CdNe and Na$^{40}$Ar (see 
Ref.~\cite{supp} for results of isotopes), reaching 
good accuracy as compared with  
experiments~\cite{kop02b,smalley79a,chan86a,tobias61a}. Quite unexpectedly, 
even for very weakly long-range bound  diatomics~\cite{w81} such as
 $^{4}$He$_{2}$, $^{40}$Ca$^{4}$He,  $^{86}$Sr$^{4}$He, and 
$^{137}$Ba$^{4}$He, we are able to find a set of  potential parameters  that 
predict a single vibrational level  at -0.107,  -67.099, -59.875, and -48.560  
$\mu$eV~\cite{supp}, consistent with the  recent literature data, -0.0999, 
-67.161, -59.573, and  -48.279 $\mu$eV, respectively~\cite{g00,lovallo04}.  

For the metastable dications He$_{2}^{++}$, Be$_{2}^{++}$,  BeH$^{++}$ and 
Mg$_{2}^{++}$, we found that they can support 5, 18, 8  and 20 vibrational 
levels that again agree with previous  studies~\cite{ornellas93a,hogreve04a}. 
Furthermore, with the new  potential function we predict that the metastable 
dication AlH$^{++}$ can  support 12 vibrational levels. The estimated lifetimes 
for the lowest four  vibrational states of BeH$^{++}$ are 
$\tau=4.9\times 10^{10}$,   
$3.3\times 10^{7}$, $4.8\times 10^{4}$,  and 130 $\mu$s, 
and those for the lowest six vibrational states of AlH$^{++}$ are 
$\tau=2.8\times 10^{16}$, $1.8\times 10^{13}$, $2.0\times 10^{10}$, 
$3.3\times 10^{7}$, $8.3\times 10^{4}$ and 288 $\mu$s (see Table 8  in 
Ref.\cite{supp}). Note that BeH$^{++}$ and AlH$^{++}$  have been recently  
observed to survive  flight times of about 4 and 7  $\mu$s, 
respectively~\cite{franzreb05a}, thus supporting our calculations. 

Before concluding we make one final mark. In the large-$R$ limit where the  
atomic electron clouds do not overlap considerably, the interaction energy  
of an atomic pair is given by the well-known multipolar dispersion  
expansion $\sum_{n=3}^{\infty}C_{2n}/R^{2n}$ \cite{kop03a,t95,herschbach70}. 
In this limit our model potential approaches $E_{\infty}$ exponentially, a 
feature different from that suggested by the multipolar dispersion expansion. 
Nevertheless, because the proposed potential is  applicable for internuclear 
distances far beyond the equilibrium position  (e.g., see Figs. 1 and 2), its 
asymptotic exponential behavior should not  present an issue except for some 
extreme cases such as ultracold collisions.  
 
In conclusion, we have proposed an analytical three-parameter potential  
function for  more than 200  weakly and strongly bound ground-state 
diatomics, including metastable molecular dications, with good 
accuracy over a significantly wide range of inter-nuclear distances. 
The determined  potential energy curves and the associated vibrational 
levels  agree well with literature data. When many-body effects are small, 
our simple pair potential function might also be useful in large-scale 
computer  simulations for complex systems. We anticipate that our model 
potential  provides a useful guide towards supplementing  the potential 
curves obtained  from the RKR and Dunham methods and a unified description 
of weakly and  strongly bound diatomics. Extensions of our molecular 
orbital theory based  approach to other types of diatomic systems are 
ongoing. 
 
This work was partially supported by DARPA and ONR. R.H.X. thanks  Garnett W.  
Bryant, Vedene H. Smith, Jr., Klaus Franzreb, Dudley R. Herschbach, Roland  
E. Allen,  Marlan O. Scully  and Hartmut Schmider for helpful discussions. 
J.G. thanks Dr. Ao Ma and Prof. Stuart Rice for reading this manuscript.


\begin{thebibliography}{99} 
 
\bibitem{g00} R. E. Grisenti, W. Sch\"{o}llkopf, J. P. Toennies, 
G. C. Hegerfeldt,  T. K\"{o}hler, and M. Stoll, Phys. Rev. Lett. 
{\bf 85}, 2284 (2000).  
 
\bibitem{w81}J. S. Winn, Acc. Chem. Res. {\bf 14}, 341 (1981). 
 
\bibitem{kop03a}J. Koperski, {\sl Van der Waals Complexes in Supersonic  
Beams} (Wiley-VCH, Weinheim, 2003).  
 
\bibitem{p01}H. Partridge, J. R. Stallcop, and E. Levin, J. Chem. Phys. 
{\bf 115}, 6471 (2001). 
 
\bibitem{t95} K. T. Tang, J. P. Toennies, and C. L. Yiu, Phys. Rev. Lett.  
{\bf 74}, 1546 (1995);  K. T. Tang, J. P. Toennies, and C. L. Yiu, 
Int. Rev. Phys. Chem. {\bf 17}, 363 (1998); K. T. Tang and  J. P. Toennies, 
J. Chem. Phys.  {\bf 118}, 4976 (2003).
 
\bibitem{k99}U. Kleinekath\"{o}fer, M. Lewerenz,  and M. Mladenovi\'{c}, 
Phys. Rev. Lett. {\bf 83}, 4717 (1999). 
 
\bibitem{e99}B. D. Esry, C. H. Greene, and J. P. Burke, Jr., Phys. Rev.  
Lett. {\bf 83}, 1751 (1999).

\bibitem{doye05} J. P. K. Doye and L. Meyer, Phys. Rev. Lett. {\bf 95}, 
063401 (2005).

\bibitem{mathur}D. Mathur, Phys. Rep. {\bf 225}, 193 (1993);  S. T. Price,  
Phys. Chem. Chem. Phys. {\bf 5}, 1717 (2003). 
 
\bibitem{kop02a} D. Steele, E. R. Lippincott, and J. T. Vanderslice,  
Rev. Mod. Phys. {\bf 34}, 239 (1962); J. Koperski, Phys. Rep. {\bf 369},  
177 (2002) 
 
\bibitem{kolos65b} W. Kolos and J. Rychlewski, J. Chem. Phys. {\bf 98}, 
3960 (1993). 
 
\bibitem{alex04a} S. A. Alexander and R. L. Coldwell, 
J. Chem. Phys. {\bf 121}, 11557 (2004). 
 
 
\bibitem{zav91a}A. A. Zavitsas, J. Am. Chem. Soc. {\bf 113}, 4755 (1991);  
H. Wei, Phys. Rev. A {\bf 42}, 2524 (1990). 
 
\bibitem{cve94}D. Cvetko, A. Lausi, A. Morgante, F. Tommasini,  
P. Cortona, and M. G. Dondi, J. Chem. Phys. {\bf 100}, 2052 (1994). 
 
\bibitem{bel02}D. Bellert and W. H. Breckenridge, Chem. Rev. {\bf 102},  
1595 (2002). 
 
\bibitem{cah04}K. Cahill and V. A. Parsegian, J. Chem. Phys. {\bf 121},  
10839 (2004). 
 
\bibitem{gill88}P. M. W. Gill, and L. Radom, Chem. Phys. Lett. {\bf 147},  
213 (1988). 
 
\bibitem{zhu} Z. H. Zhu, F. Wang, B. Cheng, M. L. Tan, and H. Y. Wang,  
Mol. Phys. {\bf 92}, 1061 (1997). F. Wang, C. Yang, and Z. H. Zhu,  
J. Mol. Struct. {\bf 684}, 9 (2004). 
 
 
\bibitem{stwalley}W. C. Stwalley, Y. H. Uang, and G. Pichler,  
Phys. Rev. Lett. {\bf 41}, 1164 (1978).  
 
\bibitem{ornellas93a} C. A. Nicolaides, M. Chrysos, and P. Valtazanos,  
J. Phys. B {\bf 23}, 791 (1990); P. J. Bruna, G. A. Di Labio, and J.  
S. Wright, J. Phys. Chem. {\bf 96}, 6269 (1992); V. V. Nefedova, A.  
I. Boldyrev, and J. Simons, Int. J. Quant. Chem. {\bf 55}, 441 (1995);   
R.H.Xie, CCSD/6-311++G(3df,3pd) results, unpublished. 
 
\bibitem{wesen} R. Wesendrup, M. Pernpointner, and P. Schwerdtfeger,  
Phys. Rev. A {\bf 60}, 3347 (1999); S. Petrie, Chem. Phys. Lett.  
{\bf 399}, 475 (2004). 
 
\bibitem{supp} See EPAPS Document No.\underline{\ \ \ \ \ \ } for 
our supporting materials. A direct link to this document may be 
found in the online article's HTML reference  section. The document  
may also be reached via the EPAPS homepage  
(http://www.aip.org/pubservs/epaps.html) or from ftp.aip.org in  
the directory /epaps/. See the EPAPS homepage for more information. 
 
\bibitem{mcquarrie}D. A. McQuarrie, \textit{Quantum Chemistry}  
(University  Science Books, Mill Valley, CA, 1983) 
 
\bibitem{peek65a} J.M. Peek, J.Chem.Phys.{\bf 43}, 3004 (1965); 
J. Patel, J. Chem. Phys. {\bf 47}, 770 (1967). 
 
\bibitem{ccsd} J. Cizek, J. Chem. Phys. {\bf 45}, 4256 (1966). 

 
\bibitem{agu94}A. Aguado, C. Suarez, and M. Paniagua, J. Chem. Phys.  
{\bf 101}, 4004 (1994). 

 
\bibitem{kop02b} J. Koperski and M. Czajkowski, Euro. Phys. J. D  
{\bf 10}, 363 (2000). 
 
 
\bibitem{smalley79a}J. Tellinghuisen, A. Ragone, M. S. Kim, D. J.  
Auerbach, R. E. Smalley, L. Wharton, and D. H. Levy, J. Chem. Phys.  
{\bf 71}, 1283 (1979). 
 
\bibitem{chan86a}Y.C. Chan, D.R. Harding, W.C. Stwalley, and C. R. Vidal,  
J. Chem. Phys. {\bf 85}, 2436 (1986). 

 
\bibitem{tobias61a}I. Tobias and J. T. Vanderslice, J. Chem. Phys.  
{\bf 35}, 1852 (1961). 
 
 
\bibitem{lovallo04} C. C. Lovallo and M. Klobukowski, J. Chem. Phys.  
{\bf 120}, 246 (2004).  
 
\bibitem{hogreve04a}H. Hogreve, Chem. Phys. Lett.  {\bf 394}, 32 (2004). 
 
\bibitem{franzreb05a}K. Franzreb, R. C. Sobers, Jr., J. L\"{o}rincik, and  
P. Williams, Phys. Rev. A {\bf 71}, 024701 (2005); Appl. Surf. Sci. {\bf 231}, 
 82 (2004).  
 
\bibitem{herschbach70}H. L. Kramer and D. R. Herschbach, J. Chem. Phys. 
{\bf 53}, 2792 (1970). 
 
\end{thebibliography}
\end{document}